\makeatletter
\declare@file@substitution{revtex4-1.cls}{revtex4-2.cls}
\makeatother

\documentclass{aastex631}

\usepackage{natbib}
\usepackage{longtable}

\usepackage{newtxtext,newtxmath}
\usepackage[T1]{fontenc}

\DeclareRobustCommand{\VAN}[3]{#2}
\let\VANthebibliography\thebibliography
\def\thebibliography{\DeclareRobustCommand{\VAN}[3]{##3}\VANthebibliography}

\usepackage{dblfloatfix}
\usepackage{graphicx}	
\usepackage{xcolor}

\usepackage{listings}
\usepackage{graphicx}                    
\usepackage{color}                       
                       
\usepackage{multirow}

\usepackage{rotating}
\usepackage{longtable}
\usepackage[normalem]{ulem}
\usepackage{xcolor}

\definecolor{refclr}{rgb}{0.56, 0.0, 1.0}

\usepackage{hyperref}
\hypersetup{colorlinks,linkcolor={blue},citecolor={blue},urlcolor={red}}

\newcommand{\fass}{   {\itshape Front. Astron. Space Sci.}}
\newcommand{\lrsp}{   {\itshape Living Rev. Sol. Phys.}}
\newcommand{\raa}{{\it Res. Astron. Astrophys.}}

\chardef\us=`\_

\begin{document}
	\title{Morphology of Solar Filaments During the Solar Maximum of Cycle 24: A Study of 35 Filaments Focusing on Chirality, Barb Orientation, Skewness, and Eruptive Behavior}			
	
	\author{Jain Jacob P.~T.}
	\affiliation{Department of Physics, National Institute of Technology Calicut, Kozhikode-673601,  India}
	\affiliation{Indian Institute of Astrophysics, Bangalore-560034,  India}
	
	\author{Ram Ajor Maurya}
	\affiliation{Department of Physics, National Institute of Technology Calicut, Kozhikode-673601, India}
	
	\begin{abstract}
		
		Filaments in the solar atmosphere are plasma structures held aloft by magnetic fields, counteracting the pull of gravity. Analyzing their magnetic structure is challenging due to the lack of direct observations of their magnetic fields.  However, through multi-wavelength intensity observations, one can infer their morphology. We identified 35 solar filaments during the maximum period of the solar cycle 24. In order to ascertain their chirality, we analyzed H$\alpha$ intensity observations. Using coronal ultraviolet intensity images, we also determined the skewness of the related coronal arcades. Furthermore, based on adjacent active regions, we classified filaments as quiescent or active, according to strength of nearby photospheric magnetic fields. We followed every filament over time to  estimates their characteristics. The hemispheric chirality preference is followed by 27 (77\,\%) filaments across all categories, based on observed bearing sense. We examined the skewness of the coronal arcades connected to 19 filaments, and they all complied with the hemisphere helicity criterion. The morphology of 16 filaments was also ascertained by integrating the skewness and bearing sense. There were two filaments with sheared arcade morphology and fourteen with flux rope shape. Additionally, by analyzing the changes in the skewness of coronal arcades and the bearing sense of filament barbs over time, we examined the variation in filament magnetic structure. Further, we examine the temporal evolution of selected filaments, which exhibit peculiar characteristics during the observational period. Lastly, we investigated the eruptive characteristics of these filaments. We did not find filaments that are more likely to explode because they do not adhere to the hemispheric chirality preference.

	\end{abstract}
	
	\section{Introduction}
	\label{s-intro}

	In the solar corona, filaments are dark, thread-like structures that are cooler and denser than the surrounding corona. These filaments, held up by the magnetic fields against the Sun's gravity,  have their origins in the chromosphere and photosphere. As they reach the solar limb due to solar rotation, they manifest as prominences, bright features visible against the the sky.  The spine of the filament represents the region of colder plasma suspended along the horizontal magnetic field. This structure is aligned  with the polarity-inversion-line (PIL), where the line-of-sight component of photospheric magnetic field vanishes. The minor polarity regions on either side of the PIL connect to the filament spine, forming filament barbs, which also appear dark. Understanding the structure and behaviour of filaments is critical, as these features play a significant role in solar activity, and studying them further could provide insights into their dynamic nature and eruptive behaviour. 
	
	Recent studies have explored the  morphology of filaments and their associated magnetic structures using both by observations and simulations. For example, \citet{Martin1994} and \citet{Martin1998} used their wire model to illustrate field lines of filament spine and barbs, while  \citet{Liewer2012} used stereoscopic techniques to  observe these field lines from multiple angles. Filaments exhibit two distinct magnetic configurations: sheared arcade, where the filament's axial magnetic field almost aligns with the photospheric field, and  flux rope, where the axial magnetic field is opposite \citep{Mackay2010}. In other words, filament has normal polarity in sheared arcade model and inverse polarity in flux rope model. Inverse arcades and normal flux ropes, which are infrequently found in the solar atmosphere, are also thought to be potential magnetic structures \citep{Gibson2018}. Understanding these magnetic configurations is essential for accurate filament classification and prediction of their eruptive behaviors. Furthermore, a sheared arcade and flux rope, which resemble normal and inverse polarity kinds, respectively, can be used to mimic a filament magnetic structure \citep{Chen2014}. High resolution magnetic field studies on coronal features or extrapolation of photospheric magnetic field are further needed to disclose specifics about their magnetic architecture.  Determining filament properties is hampered by the high computing cost of extrapolation and sporadic observations of the coronal magnetic field.
	
	Simulation have played a pivotal role in interpreting the formation and magnetic structure of filaments. \citet{DeVore2000} demonstrated that sheared magnetic arcades form as a result of shearing motions at the photospheric footpoints, especially near the polarity inversion line (PIL). These sheared fields lead to  magnetic flux cancellations, which contribute to the formation of helical magnetic field lines, forming the foundation of the flux rope model \citep{Ballegooijen1989}. According to this model, the plasma in filaments resides in the dips of these helical magnetic field lines, and the  magnetic field within the filament is of inverse polarity compared to the surrounding photospheric fields \citep{Mackay2010}. Such simulations provide valuable insights into the underlying magnetic structures that govern filament dynamics, but additional research is needed to refine these models and connect them with observational data.

	Filaments are typically classified based on morphology, dynamics, and location into two groups, quiescent prominence and active prominence~\citep{Labrosse2010}. Quiescent prominences, found at higher latitudes, are stable and can persist for weeks to months. On the other hand, active region prominences, located near sunspots, are short-lived and are more prone to eruptions. Intermediate filaments bridge the gap between quiescent and active types, forming between weak background fields and active regions. Quiescent prominences can also  undergo eruptions due to instabilities, leading to their classification into eruptive and non-eruptive types. Moreover, the filaments are categorized based on their magnetic properties, such as helicity and chirality. The study of filament chirality can offer valuable insights into the filament's axial field direction and magnetic helicity, both of which are essential to understanding filament dynamics and eruptions.

	However, determining the magnetic helicity of filaments remains challenge due to the lack of magnetic field observations above the photosphere. The chirality of a filament is related to underlying magnetic structures. It can provide insights into the filament’s axial field direction and magnetic helicity \citep{Hao2016}. Filament chirality of majority of filaments can be easily determined using plenty of observations on chromospheric and coronal intensities. Researchers has developed various techniques for chirality determination. \citet{Martin1994} and \citet{Martin1998} classified filaments as dextral or sinistral based on the direction of their axial field relative to the positive polarity side and the orientation of their barbs. A filament is considered dextral if the axial field points rightward when viewed from the positive polarity side, while sinistral filaments have axial fields directed leftward. Barbs are classified as right-bearing or left-bearing based on their veering direction from the filament spine when viewed from either footpoint. This method links chirality to magnetic helicity, where dextral filaments correspond to negative helicity and sinistral filaments to positive helicity.  
	
	Further, \citet{Pevtsov2003} determined filament chirality quantitatively by measuring fractional chirality based on the orientation of individual barbs in H-alpha images. Similarly, \citet{Hazra2018} estimated chirality by counting the number of right-bearing and left-bearing barbs. This method allows chirality classification based on visual inspection of barb orientation without needing magnetic field data. However, \citet{Guo2010} identified a filament with mixed left- and right-bearing barbs along its axis. The existence of flux rope and sheared arcade present in different regions of filament magnetic structure leads to the presence of opposite bearing sense in the same filament.	However, the correlation between morphology and magnetic helicity is not always straightforward. \citet{Aparna2024} found that only 35\% of filaments in their study matched the predicted relationship between chirality, magnetic helicity, and force-free parameters. This discrepancy suggests that additional information about filament magnetic structure is needed to accurately determine magnetic helicity.
	
	In addition to chirality, the combined observations on skewness of coronal arcades associated with filament and its bearing sense  also contribute in chirality estimation. \citet{Martin1998} used the skew of magnetic loops around the filament to assess the chirality of coronal arcades that lie on top of filaments. Left-skewed loops relative to the filament axis correspond to dextral filaments, while right-skewed loops indicate sinistral filaments. These findings were confirmed by \citet{Joshi2014}, who observed left-bearing barbs in conjunction with right-skewed arcades, consistent with sinistral chirality. Furthermore, studies such as \citet{Liewer2012} determined sinistral chirality in another filament combining observations on barb orientation and magnetograms. Also, they found prominence rotates clockwise during the eruption, as reported in \citet{Tripati2011}. \textit{i.e.,} the direction of prominence rotation during eruptions can indicate their chirality \citep{Zhou2020}. These combined techniques, integrating both visual and magnetic data, allow for a more precise classification of filament chirality and its relation to magnetic helicity.
	
	Moreover, the previous studies \citep{Guo2010,Chen2014,Chen2020} have revealed that the orientation of filament barbs is influenced by the underlying magnetic structures. For instance, dextral filaments with right-bearing barbs are associated with inverse-polarity filaments, while left-bearing barbs tend correspond to normal-polarity configurations. According to \citet{Chen2014}, magnetic configuration of a inverse-polarity filament was described by flux rope model and normal-polarity filament by sheared arcade model. In solar atmosphere, a prevalence of inverse-polarity filaments were observed \citep{Leroy1984}. This leads to a hemisphere trend in chirality, with right-bearing barbs predominantly found in the Northern Hemisphere and left-bearing barbs in the Southern. The earlier study on filament chirality by \citet{Pevtsov2003} defined dextral filaments as filaments following the right-bearing sense and sinistral filaments as filaments following the left-bearing sense. However, these findings may be influenced by the inverse-polarity configuration that appears in most filaments. Therefore, determining filament chirality based purely on barb orientation can lead to erroneous conclusions without considering the underlying magnetic structure.

	To better understand the  magnetic structure of filaments, some studies have used extrapolated photospheric magnetic field \citet[e.g.,][]{Guo2010}, while others have modelled filament structures using MHD simulations \citep{DeVore2000,Fan2019}.   \citet{Chen2020}, however, proposed a method for estimating filament magnetic structure using observations of both the skewness of coronal arcades and the bearing sense of barbs together, which does not rely on computationally expensive numerical techniques. Given the weak magnetic fields in the quiet sun region, it is often challenging to extrapolate magnetic fields accurately. Thus, this study aims to use this method to study the magnetic structure of filaments.
	
	Filaments are dynamic structures that undergo changes in their magnetic configuration and physical properties throughout their lifecycle. Continuous monitoring and analysis of filaments are essential to uncover the full extent of their characteristics. In this study, we estimate the chirality of filaments at various stages of their evolution and track their consistency over time. Additionally, we categorize filaments based on their associated active regions, eruptive behavior, magnetic structure, and chirality. This multi-faceted approach will provide a clearer understanding of filament behavior and its implications for solar activity.   
	
	The rest of the paper is organized as follows: Section~\ref{s-obs} presents a detailed description of the data used in this study, including the sources and methods of data collection. Section~\ref{s-analysis} outlines the analysis process and discusses the results obtained from the study. Finally, Section~\ref{s-summary} provides a summary of the key findings and concludes with the implications of the results, highlighting potential avenues for future research.
	
	\section{Observational data}
	\label{s-obs}

	To analyze the filament chirality, we utilized data from two ground based observatories: the Kanzelh$\rm \ddot{o}$ehe Solar Observatory (KSO) in Austria and the Global Oscillation Network Group \citep[GONG,][]{Harvey1996}  at Cerro Tololo Inter-American Observatory (CT) in Chile. Both observatories provide full-disk H-alpha images with pixel resolutions of 1\arcsec.02 and 1\arcsec.0, respectively. These H-alpha images were instrumental in estimating the bearing sense of the solar filaments. Our analysis focused on filaments observed from December 1, 2013 to January 31, 2014 providing an adequate dataset for examining filament morphology during this period. 
	
	In addition to the ground-based H-alpha observations, we determined the  skewness of associated coronal loops using the extreme ultraviolet (EUV) intensity data collected during the same observation period.  For this, we relied on data from the Atmospheric Imaging Assembly  \citep[AIA,][]{Lemen2011} onboard Solar Dynamics Observatory that offers full-disk solar observations with a time cadence of 12 seconds. Solar filaments are observed in the SDO/AIA in EUV wavelengths of 171\,\AA, 193\,\AA\ and 304\,\AA\ with a pixel resolution of 0\arcsec.6. The coronal loops above the filaments are observed also in EUV wavelengths of 171\,\AA, 193\,\AA\ and 211\,\AA. The examination of the coronal loops above the filaments enhance our understanding of their structure and orientation.
	
	To identify the presence of active regions associated with filaments and study their magnetic fields, we incorporated line-of-sight magnetic field data from the Helioseismic and Magnetic Imager (HMI) onboard SDO. The SDO/HMI provides magnetic field data with a pixel resolution of  0\arcsec.6 and a temporal resolution of 45\,seconds. This data is crucial for analyzing the magnetic environment surrounding the filaments and their associated active regions, offering valuable insights into the relationship between filament structure and magnetic activity in the solar atmosphere.

	\section{Analysis and Results}
	\label{s-analysis} 
	
	We studied the properties of 35 solar filaments observed from  December 1, 2013 to January 31, 2014. Initially, these filaments were identified and labeled as FL01 through FL55. We tracked their evolution through multiple observations and recorded their changes over time. As we continued to track the filaments, we identified repeated observations and reassigned each filament a new label, FLM01 through FLM35, to reflect their updated status.
	
	To better organize the filaments, we classified them into different categories
	based on their observed characteristics. This classification was essential for understanding the variations in filament morphology and dynamics across the observation period. Additionally, we estimated chirality and magnetic structure of each filament, providing crucial insights into their underlying magnetic properties and their role in solar activities. 
	
\begin{figure}[t]
		\centering
		\includegraphics[width=0.8\textwidth,clip,viewport=8 267 605 788]{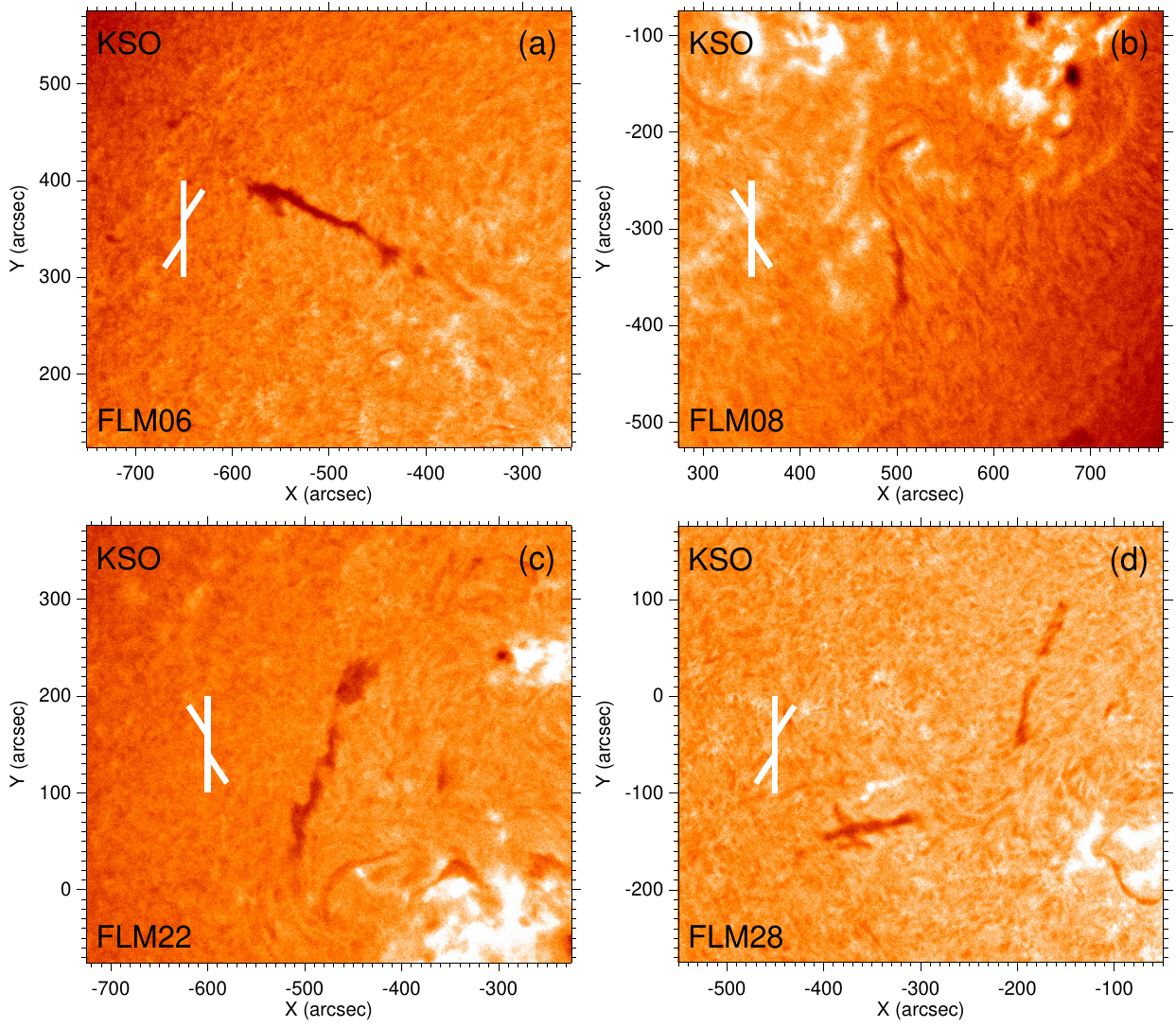}
		\caption{H-alpha intensity images of four filaments observed by KSO, with the white crosses in each panel representing the filament orientation.}
		\label{fig:fil-halpha-barb}
	\end{figure}

\begin{figure}[h]
		\centering
		\includegraphics[width=0.8\textwidth,clip,viewport=10  267 604 790]{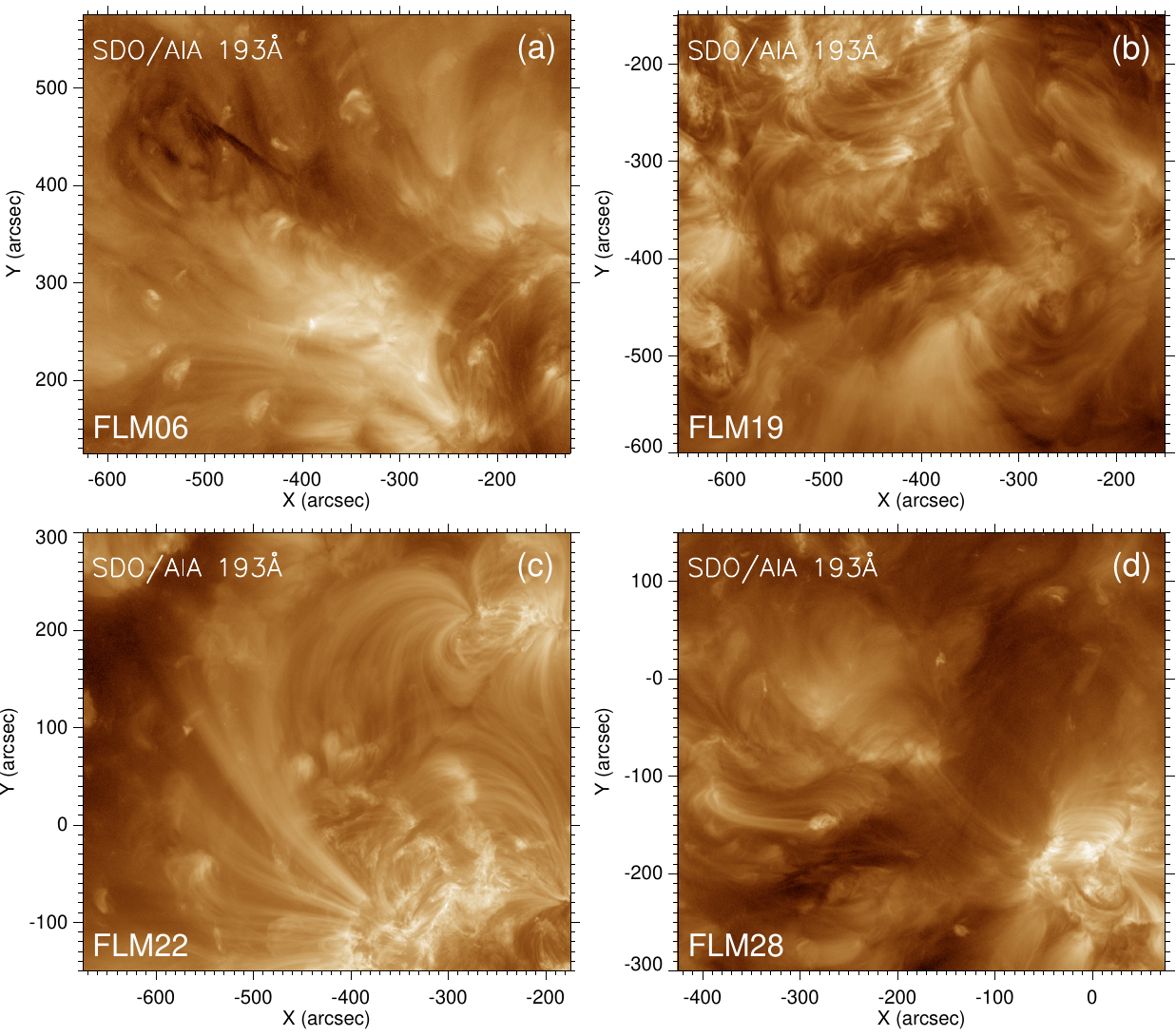}
		\caption{The SDO/AIA 193\,\AA\ intensity images of coronal arcade associated with filament (a) FLM06, (b) FLM19, (c) FLM22 and (d) FLM28 as shown in Figure~\ref{fig:fil-halpha-barb}.}
		\label{fig:fil-arcade-skew}
	\end{figure}

	The results of our analysis, including detailed observations and categorization, are presented in Figures~\ref{fig:fil-halpha-barb} to ~\ref{fig:flm28-gong-aia}. A summary of the filament properties, including their morphology, chirality and magnetic characteristic, is provided in Tables~\ref{tab-filaments-morph} to~\ref{tab-summary-table}. These results offer a comprehensive view of the filaments' behaviour and structure during the study period.

	\subsection{Bearing sense of Chromospheric filaments}
	\label{ss-fil-chir} 
	
	We determined the bearing sense of  filaments using the method outlined by \cite{Hanaoka2017}. The left- and right-bearing orientation of a filament barb can be understood as follows: initially, move along the filament spine in a direction to make acute angle with a specific barb from direction of motion. If this barb is observed on right side of the filament spine, it is considered as right-bearing. Conversely, if this barb is observed on left side of the filament spine, it is considered as left-bearing. The H-alpha images of filaments FLM06 and FLM28 with right bearing barbs are shown in Figure~\ref{fig:fil-halpha-barb}(a) and (d), observed in northern and southern hemisphere, respectively. We overlaid a schematic representation of their bearing sense in each panels. Similarly, in Figure~\ref{fig:fil-halpha-barb}(b) and (c) we displayed H-alpha images of filaments FLM08 and FLM22, which exhibit left bearing barbs.
	
	Further, we calculated fractional chirality ($C_{\rm f}$) by considering the  number of right- and left-bearing barbs, denoted as $N_{\rm R}$ and $N_{\rm L}$, respectively. For this, we examined H-alpha images captured by GONG/CT and KSO. According to \citet{Pevtsov2003}, the fractional chirality is given by the equation,
	
	\begin{equation}
		C_f=\frac{N_{\rm R}-N_{\rm L}}{N_{\rm R}+N_{\rm L}}.
		\label{fra-chirality-equ}
	\end{equation}
	
	\noindent The detected bearing orientation and $C_{\rm f}$ value for each filament are listed in Table~\ref{tab-filaments-morph}. Ambiguous cases, footnoted and excluded from the summary, highlight challenges in consistent classification due to factors like low $|C_f|$, changes in bearing/skewness, projection effects, and unclear observations. The fractional chirality ($C_f$) values range from -1.00 to 1.00, indicating their bearing of barbs and senses of twist. 
	
	Our findings indicate that filaments with right-bearing barbs are predominantly located in the northern hemisphere, while left-bearing barbs are more commonly observed  in the southern hemisphere. In total, 46\% of filaments exhibit left-bearing sense in their barbs, while 40\% show right-bearing sense. The remaining 14\% of filaments have erroneous chirality values.

	\subsection{Skewness of Coronal Arcade}
	\label{ss-arc-skew}

	We estimated the skewness of coronal arcades, which are loop structures above solar filaments. Skewness is defined in relation to the filament axis and the direction of the filament barbs, as described by \citep{Martin1998}. When viewed from above, right-skewed arcades form an angle with the filament axis in the same direction as the right-bearing barbs. Similarly, left-skewed arcades form an angle with the filament axis that aligns with the left-bearing barbs. Left-skewed arcades are typically found above dextral (left-handed) filaments, while right-skewed arcades are found over sinistral (right-handed) filaments.

	To analyze the skewness of coronal arcades, we used EUV intensity images from the SDO/AIA observations at wavelengths of 171\,\AA, 193\,\AA, and 211\,\AA. Figure~\ref{fig:fil-arcade-skew} shows SDO/AIA 193\,\AA\ images of coronal arcades in the region of four filaments. From these EUV images, we found that the filaments FLM06, FLM22 and FLM28 exhibited left-skewed arcades, while FLM19 displayed a right-skewed arcade. In total, we identified the skewness of  coronal arcades corresponding for 20 filaments, as summarized in Table~\ref{tab-filaments-morph}. However, the skewness of coronal arcade in one filament could not be determined due to observed change in skewness. The remaining 19 filaments adhered to the hemispheric helicity rule based on the skewness of their associated arcades. 
	
	The skewness of the coronal arcades associated with filaments varies. 46\% of filaments exhibit a right-skewed arcade, and 8\% show a left-skewed arcade. However, for nearly half of the filaments (46\%), skewness data is not available. We determined magnetic structure of 46\% of filaments with clear observations on their skewness of overlying coronal arcades. By considering filaments whose skewness are determined, the majority of filaments (87\%) are associated with flux rope morphology. Sheared arcade morphology is observed in 13\% of the filaments.
	
	\begin{figure}[b]
		\centering
		\includegraphics[width=0.8\textwidth,clip,viewport=30 311 605 788]{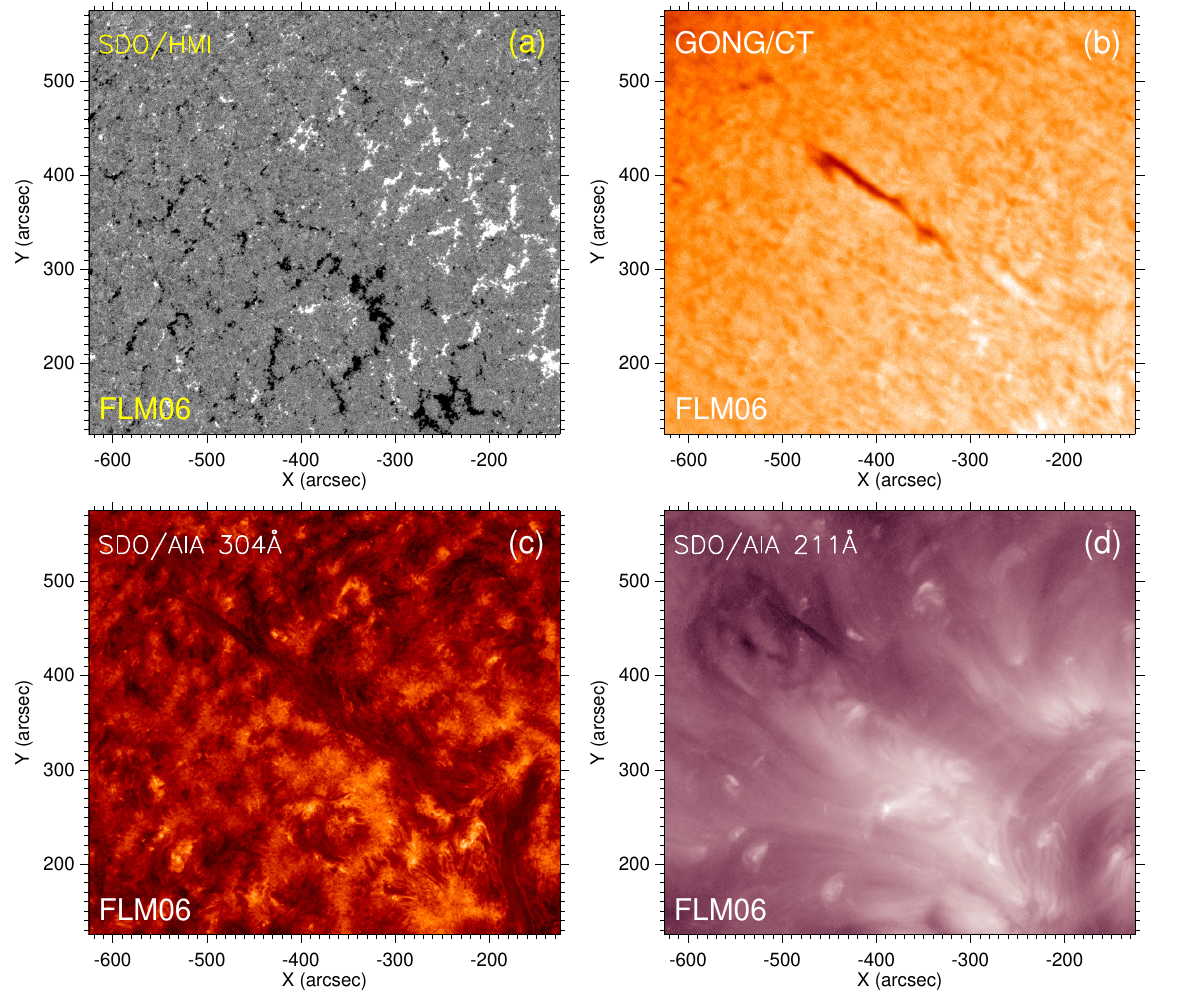}
		\caption{(a) LOS magnetogram (b) H-alpha image (c) SDO/AIA 304\,\AA\ image (d) SDO/AIA 211\,\AA\ image of filament tagged as FLM06.}
		\label{fig:fl8-mag-chromo-arcade}
	\end{figure}

	\begin{figure}[h]
		\centering
		\includegraphics[width=0.8\textwidth,clip,viewport=10 267 611 790]{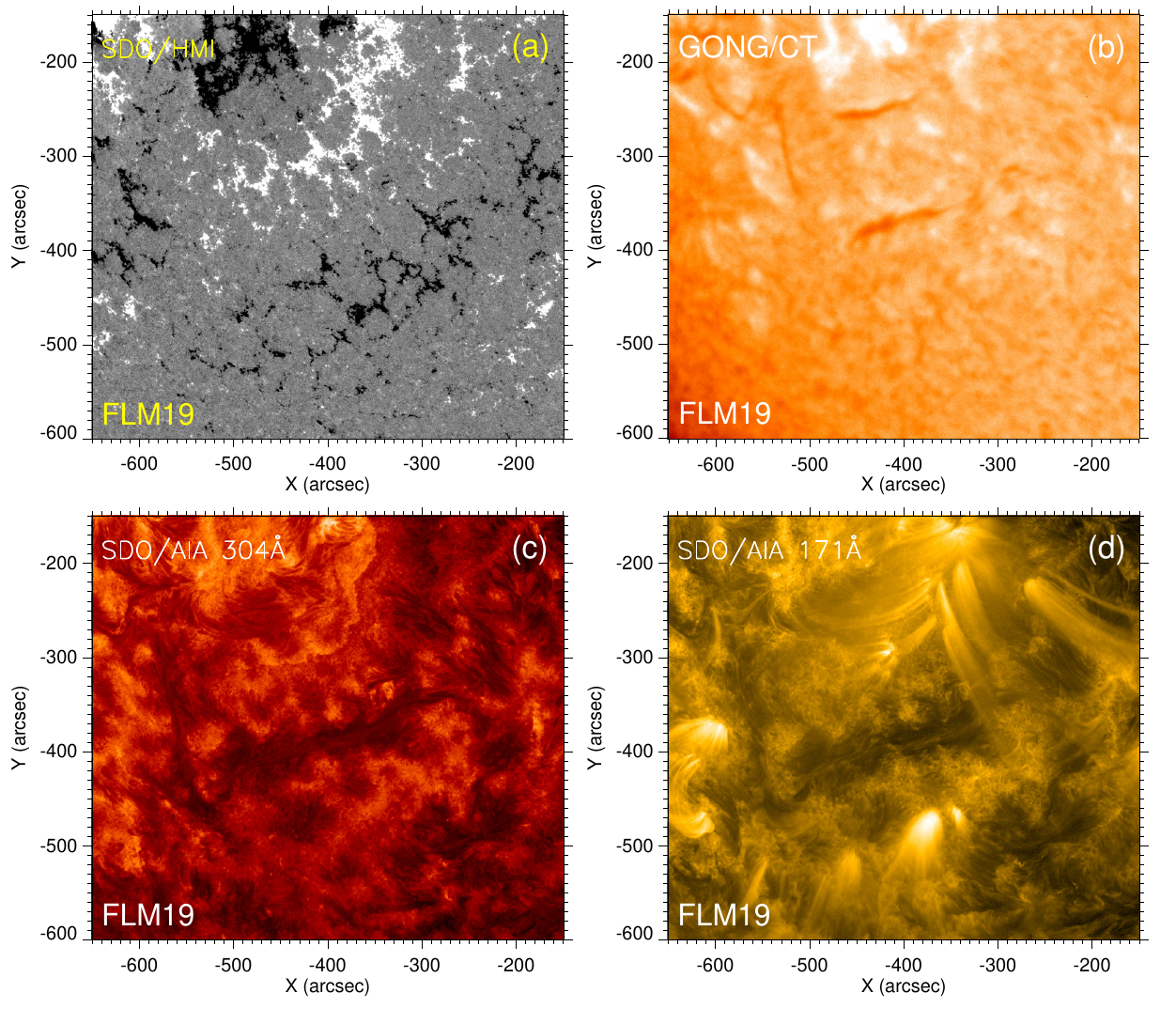}
		\caption{(a) LOS magnetogram (b) H-alpha image (c) SDO/AIA 304\,\AA\ image (d) SDO/AIA 171\,\AA\ image of filament tagged as FLM19.}
		\label{fig:fl22-mag-chromo-arcade}
	\end{figure}  
	
	\begin{figure}[h]
		\centering
		\includegraphics[width=0.8\textwidth,clip, viewport=14 280 611 790]{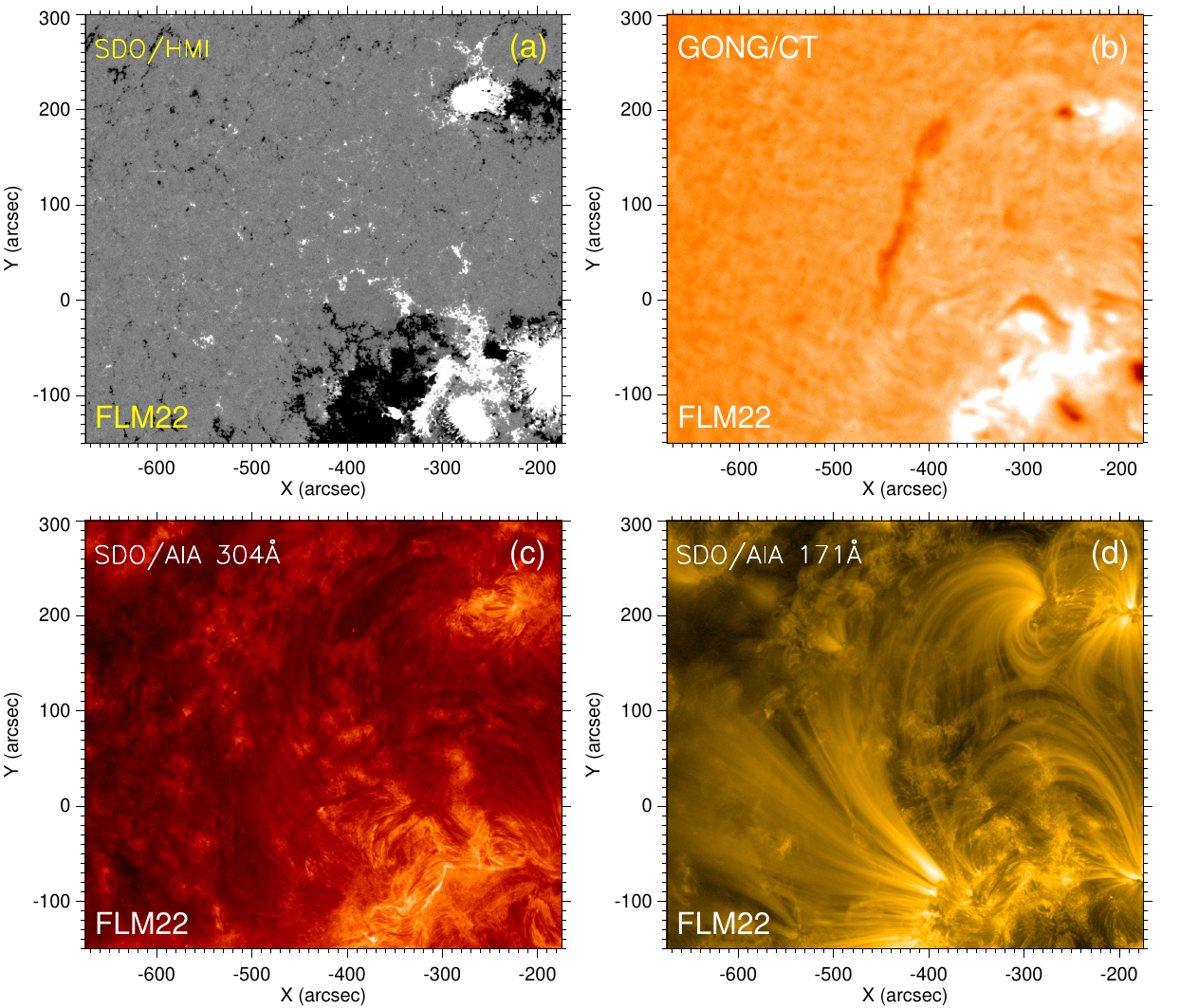}
		\caption{Line-of-sight magnetic field (a), H-alpha intensity (b), and EUV intensities in 304\,\AA\ (c) and 171\,\AA\ (d) observations of the filament FLM22.}
		\label{fig:fl27-mag-chromo-arcade}
	\end{figure}

	\subsection{Classification of Chromospheric Filaments}
	\label{ss-fil-class}
	
	To identify the presence of active regions near solar filaments, we examined magnetograms from SDO/HMI and Helioviewer. Based on these observations, we identified 10 filaments associated with active regions, which were classified as active region filaments (labeled as AR). Another 10 filaments were categorized as quiescent filaments (labeled as QR), as they were observed in quiet regions with no sunspots. The remaining 15 filaments were intermediate filaments (labeled as IF) as they were located near relatively weak active regions. 
	
	In Figure~\ref{fig:fl8-mag-chromo-arcade}, we present the observations of the filament FLM06 with a line-of-sight magnetogram (panel a), H-alpha intensity (panel b), and EUV intensities (panels c and d) in different wavelengths. This filament is classified as a quiescent filament according to SDO/HMI line-of-sight magnetogram in the panel (a). We found three right-bearing barbs in H-alpha image as shown in panel (b). The filament plasma and its associated coronal arcade are clearly visible in SDO/AIA 304\,\AA\ and SDO/AIA 211\,\AA, as shown in panels (c) and (d), respectively. We observed left-skewed coronal arcade associated with this filament and classified as dextral filament. Additionally, the observed line-of-sight magnetic field, H-alpha image and EUV images in different wavelengths of other quiescent filament FLM19 is shown in Figure~\ref{fig:fl22-mag-chromo-arcade}. The FLM19 has five left-bearing barbs and right-skewed coronal arcade as per the observations. It is classified into sinistral filament. On the other hand, an active region filament named FLM22 is shown in Figure~\ref{fig:fl27-mag-chromo-arcade}. We found an active region at southern foot point of this filament as shown in panel (a) and four left-bearing barbs in H-alpha image in panel (b). The filament plasma and associated coronal arcade were clearly observed in SDO/AIA 304\,\AA\ and SDO/AIA 171\,\AA, as shown in panel (c) and (d), respectively. We observed left-skewed coronal arcade associated with this filament and classified into dextral filament. Similarly, we determined chirality of all filaments and details of observations are tabulated in Table~\ref{tab-filaments-morph}.
	
	Throughout the observation period, we noted that some filaments underwent eruptions, prompting us to  classify them as either eruptive or non-eruptive. All quiescent filaments were found to be non-eruptive. However, we detected one eruptive active region filament and nine eruptive intermediate filaments in our study. Regarding eruption status, the majority of filaments (71\%) are non-eruptive, meaning no eruption was observed during the observation period. Eruptive filaments, which are associated with solar eruptions, comprise 28\% of the dataset.
	
	In addition to this, we investigated  the magnetic structure of filaments, as discussed in Section~\ref{ss-fil-stru}. Based on this analysis, we classified the filaments into two categories: sheared arcade and flux rope. This classification of filaments are also listed in the Table~\ref{tab-filaments-morph}.

	\subsection{Magnetic Structure of Chromospheric Filaments}
	\label{ss-fil-stru}
	
	We determined the magnetic structure of these filaments based on simultaneous observations on their bearing sense and skewness of associated coronal arcade, as described by \citet{Chen2020}. Filaments with left-skewed arcades and right-bearing barbs align with the flux rope morphology, while those with left-skewed arcades and left-bearing barbs   correspond to a sheared arcade morphology. Conversely, right-skewed arcades with left-bearing barbs indicate flux rope morphology, while right-skewed arcade with right-bearing barbs suggest sheared arcade morphology.

	To ensure an accurate determination of the filaments morphology, we analyzed each filament multiple times. We observed changes in both the bearing sense of barbs and the skewness of the coronal arcades during the evolution of some filaments. In few cases, the bearing sense of barbs was unclear, leading to exclusion of these filaments from their magnetic structure analysis.
	
	By combining observations of skewness and bearing sense, we accurately determined the morphology of 16 filaments. For example, the Figure~\ref{fig:fl8-mag-chromo-arcade} shows the skewness of coronal arcade and bearing sense of barbs in FLM06. It exhibits right-bearing barbs and left skewed arcade, which, according to \citet{Chen2020}, corresponds to flux rope morphology. In contrast, the filament FLM22 shown in Figure~\ref{fig:fl27-mag-chromo-arcade}, displays left-bearing barbs and a left skewed arcade, classifying it as a sheared arcade morphology. We find that 14 of the 16 filaments exhibited flux rope morphology, while the remaining two exhibited sheared arcade morphology. Notably, all quiescent filaments followed the flux rope morphology. 
	
\begin{figure}[h]
		\centering
		\includegraphics[width=\textwidth,clip,viewport=210 133 2748 1322]{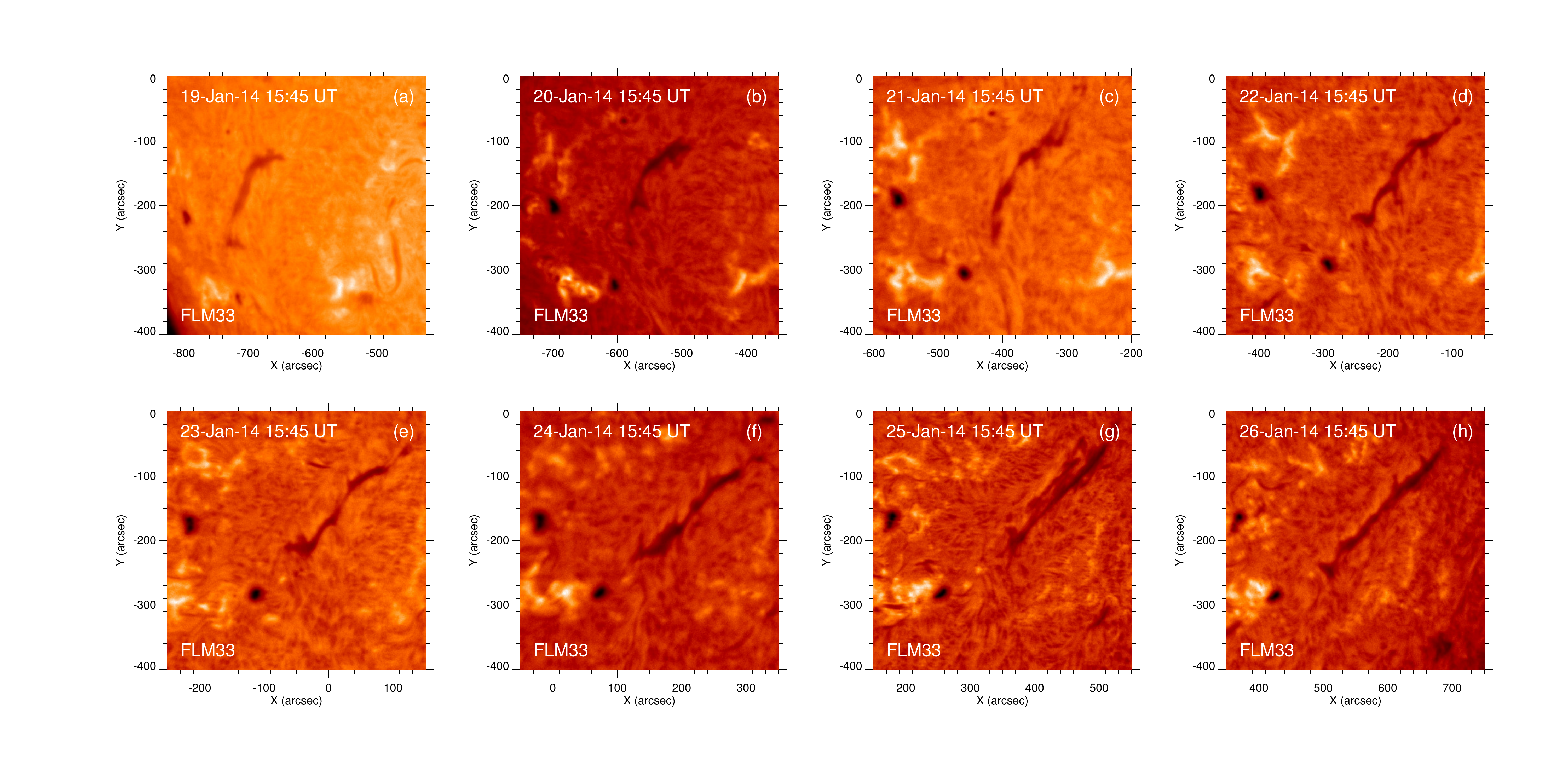}
		\caption{The GONG/CT H-alpha observations on bearing sense of barbs on filament FLM33 during its evolution from January 19, 2014 to January 26, 2014.}
		\label{fig:flm33_evol-gong}
	\end{figure}

\begin{figure}[h]
		\centering
		\includegraphics[width=\textwidth,clip,viewport=210 133 2748 1322]{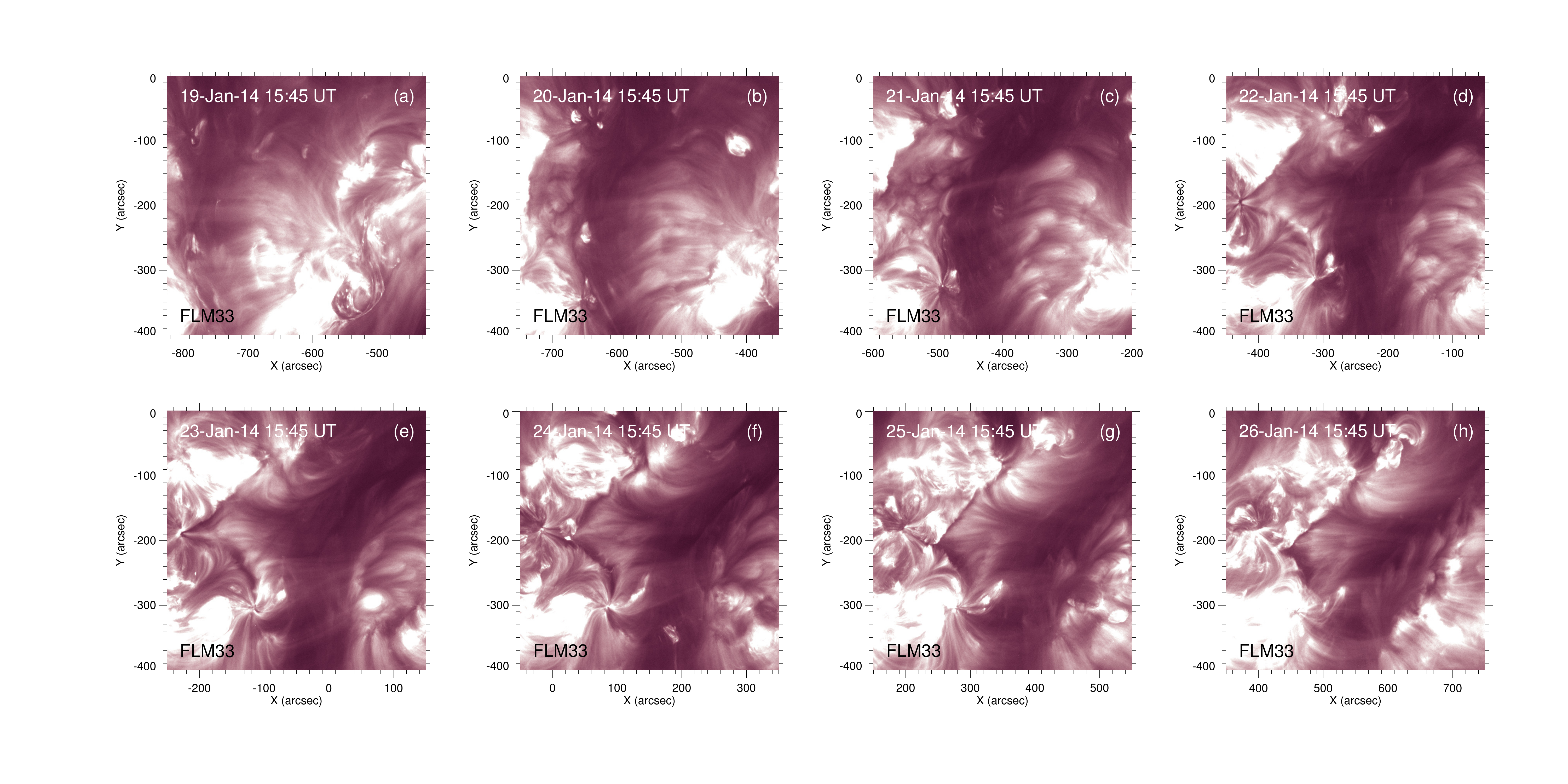}
		\caption{The SDO/AIA 211\,\AA\ observations on skewness of coronal arcade above the filament FLM33 during its evolution from January 19, 2014 to January 26, 2014.}
		\label{fig:flm33-evol-aia}
	\end{figure}

	In a related study by \citet{JainJacob2025a}, the filament FLM33 exhibited initially a right bearing sense. The evolution of filament FLM33 is shown in Figure~\ref{fig:flm33_evol-gong} and~\ref{fig:flm33-evol-aia}. The filament FLM33 has two right bearing barbs at a location \mbox{(-500\,arcsec, -125\,arcsec)} on January 20, 2014 as shown in Figure~\ref{fig:flm33_evol-gong}(b). Although the January 19, 2014 observations are affected by projection effects due to the filament's proximity to the solar limb, the observed morphology is consistent with an initial right-bearing interpretation (Figure~\ref{fig:flm33_evol-gong}(a)). However, as seen in Figure~\ref{fig:flm33_evol-gong}(b), when barbs are oriented nearly perpendicular to the filament spine, manually determining their bearing from a single H$\alpha$ filtergram can be ambiguous and subject to different visual interpretations by different observers. The temporal evolution of the filament prior to its transition to a left-bearing configuration provides additional observational context supporting our interpretation.
	 On 21 January, FLM33 exhibited a transition from a right-bearing to a left-bearing configuration, where projection effects are minimal. This change in the bearing sense is summarized in Table~\ref{tab-filaments-morph}. Such a change is consistent with the predominance of left-bearing filaments in the southern hemisphere. The filament is associated with active region NOAA 11959, where an enhancement in positive helicity has been reported in a recent study \citet{JainJacob2025a}. The emergence of positive helicity in the active region located at the filament footpoint is likely to influence the filament’s evolution, leading it to conform to the hemispheric helicity rule. In this paper, we further examined the skewness of the coronal arcade above this filament. The observed bearing sense of filament barbs and associated coronal arcades of filament FLM33 from January 19 to January 25 are mentioned in Table~\ref{tab-filaments-morph}. These combined observations suggest that the filament underwent a magnetic structural transition from a sheared arcade to a flux rope. The evolution into a flux rope configuration is accompanied by the emergence of positive helicity from the photosphere, which is consistent with the dominant helicity in the southern hemisphere during the later stages. Throughout the observational period, the filament remained stable in the solar atmosphere. Our results indicate that the filament not only transforms into a flux rope morphology but also conforms to the hemispheric helicity rule, maintaining long-term stability in the solar atmosphere.
	
\begin{figure}[ht]
		\centering
		\includegraphics[width=0.8\textwidth,clip,viewport=1 276 611 790]{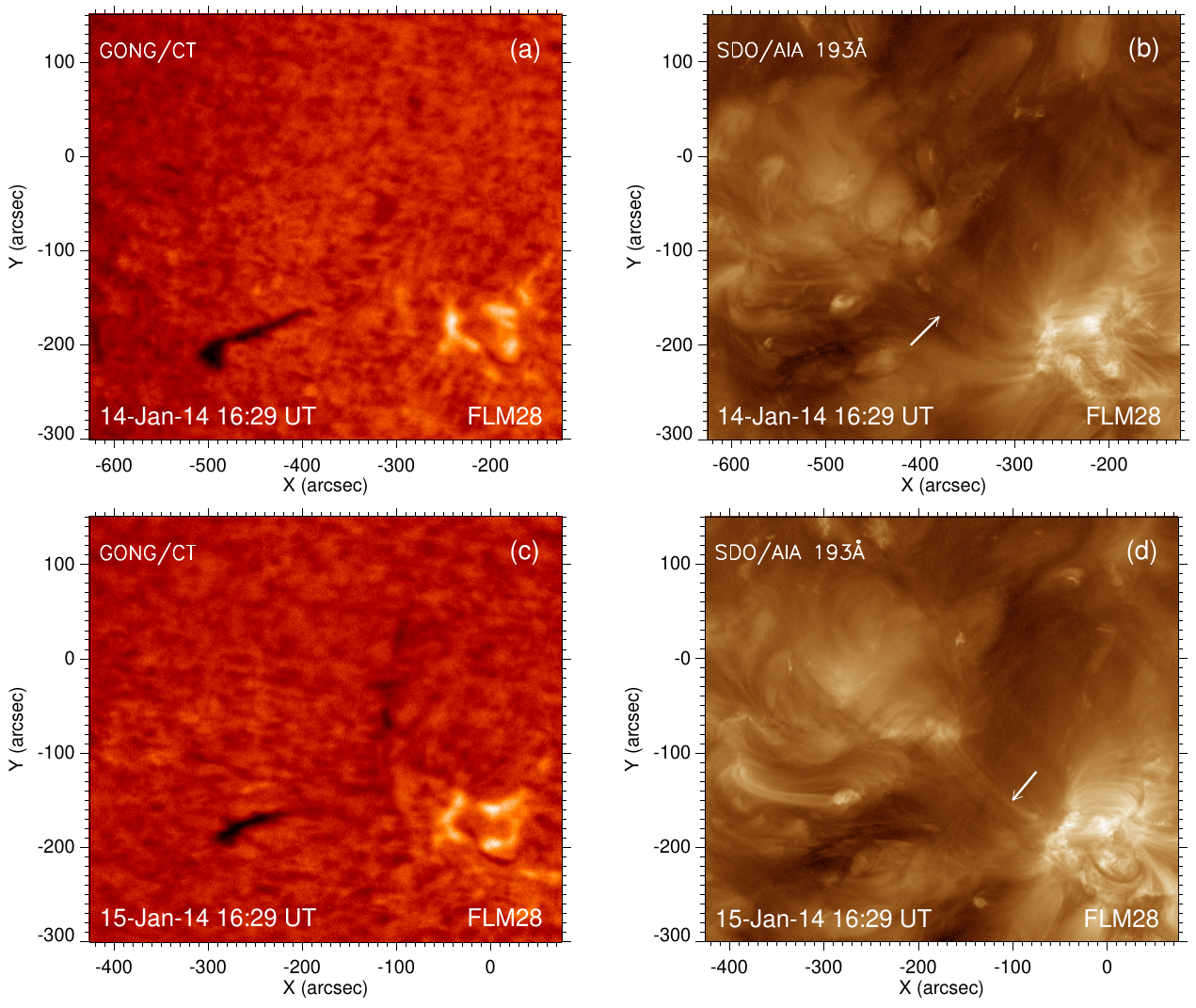}
		\caption{(a) GONG/CT H-alpha image and (b) SDO/AIA 193\,\AA\ image of filament FLM28 on January 14, 2014. (c) GONG/CT H-alpha image and (d) SDO/AIA 193\,\AA\ image of same filament on January 15, 2014. The observed coronal arcades above the filaments are marked with white arrows in SDO/AIA 193\,\AA\ images. (Animation of the filament evolution in SDO/AIA wavelength 193\,\AA\ is available in the online version.)}
		\label{fig:flm28-gong-aia}
	\end{figure}

	The bearing sense of the barbs and the skewness of the coronal arcades associated with FLM28 are shown in Figure~\ref{fig:flm28-gong-aia}. The filament consistently exhibits right-bearing barbs throughout the observations on January 13 and 14 (see, Figure~\ref{fig:flm28-gong-aia}(a)), and on 15 (see, Figure~\ref{fig:flm28-gong-aia}(c)). In contrast, the overlying coronal arcade displays a changing skewness pattern, appearing left-skewed during the first observation on January 13, right-skewed on January 14 (marked by a white arrow in Figure~\ref{fig:flm28-gong-aia}(b)), and again left-skewed on January 15 (marked by a white arrow in Figure~\ref{fig:flm28-gong-aia}(d)). Owing to the relatively low fractional chirality present in the first observation on January 13, the inferred transition from a flux-rope to a sheared-arcade configuration remains uncertain. However, the subsequent evolution from a sheared arcade to a flux rope is more convincingly supported by the observed reversal in coronal arcade skewness. This filament, exhibiting a dynamically changing magnetic configuration, ultimately becomes eruptive.

	The detailed morphology of all filaments is presented in Table~\ref{tab-filaments-morph}. The table catalogs each filament (FLM No. and FL. No.) by date and heliospheric location (latitude and longitude). Morphological characteristics include filament type (IF, AR, QR), eruptive nature (E/N), bearing sense of barbs (R/L), fractional chirality ($C_f$), skewness of the associated coronal arcade, and the magnetic structure.

	\renewcommand{\arraystretch}{1.3}
	\begin{longtable}{|cccccccccc|}
		\caption{Details of filaments' morphology: filament type (F.T.), erruptive or non erruptive (E/N), right or left bearing (R/L), fractional chirality ($C_{\rm f}$) and skewness of the associated coronal arcade. Ambiguous results are mentioned with superscripts 1--5 and expansion of superscripts are mentioned in footnote.}\label{tab-filaments-morph}\\
		
		\hline
		
		\textbf{FLM. No.} & \textbf{FL. No.} & \textbf{Date} &\textbf{ Location} & \textbf{F.T.} & \textbf{E/N} &\textbf{ R/L} & $C_f$ & \textbf{Skew.} & \textbf{Structure of Filament} \\
		\hline
		\endfirsthead
		\hline
		
				\textbf{FLM. No.} & \textbf{FL. No.} & \textbf{Date} &\textbf{ Location} & \textbf{F.T.} & \textbf{E/N} &\textbf{ R/L} & $C_f$ & \textbf{Skew.} & \textbf{Structure of Filament} \\
				\hline
				\endhead
				
				\hline \multicolumn{10}{|r|}{{Continued on next page}} \\ \hline
				\endfoot
				\hline \hline
				\endlastfoot
				
				FLM01 & 1 & 2013/12/01 & (19N,35E) & IF & E & R & 0.50 &  &  \\
				\hline
				FLM02 & 2 & 2013/12/02 & (23S,42E) & AR &  & L & -1.00 & right & Flux rope \\
				\hline
				FLM03 & 3 & 2013/12/08 & (29N,3W) & QR &  & R & 0.50 &  & \\
				\hline
				FLM04 & 4 & 2013/12/12 & (22N,33E) & QR &  & R & 1.00 &  &  \\
				\hline
				FLM05 & 5 & 2013/12/12 & (10N,47W) & AR &  & R & 1.00 &  & \\
				\hline
				FLM06$^4$ & 6 & 2013/12/14 & (18N,46E) & QR &  & L & -1.00 &  &  \\
				\hline
				FLM07$^{1,2}$ & 7 & 2013/12/14 & (28S,51E) & QR & E & L & -0.33 &  &  \\
				\hline
				FLM06 & 8 & 2013/12/16 & (19N,25E) & QR &  & R & 1.00 & left & Flux rope \\
				\hline
				FLM07$^{2}$ & 9 & 2013/12/16 & (28S,27E) & IF & E & R & 1.00 &  &  \\
				\hline
				FLM08 & 10 & 2013/12/16 & (26S,29W) & IF &  & L & -1.00 & right & Flux rope \\
				\hline
				FLM09 & 11 & 2013/12/24 & (17S,50W) & AR &  & R & 1.00 & right & Sheared Arcade \\
				\hline
				FLM10 & 12 & 2013/12/27 & (35N,27E) & QR &  & R & 1.00 &  &  \\
				\hline
				FLM11$^{1,2}$ & 13 & 2013/12/28 & (1N,47E) & AR & E & R & 0.33 &  &  \\
				\hline
				FLM12 & 14 & 2013/12/29 & (24S,44W) & AR &  & L & -1.00 & right & Flux rope \\
				\hline
				FLM13$^{1}$ & 15 & 2013/12/31 & (33S,3W) & IF & E &  & 0.00 & right & NA \\
				\hline
				FLM14 & 16 & 2014/01/01 & (29N,18E) & QR &  & R & 1.00 &  &  \\
				\hline
				FLM15 & 17 & 2014/01/01 & (13N,40W) & QR &  & R & 0.50 &  &  \\
				\hline
				FLM16 & 18 & 2014/01/01 & (11N,34E) & IF &  & R & 1.00 &  &  \\
				\hline
				FLM11$^{2}$ & 19 & 2014/01/02 & (2N,16W) & IF & E & L & -1.00 &  &  \\
				\hline
				FLM17 & 20 & 2014/01/02 & (7S,12W) & IF &  & L & -1.00 & right & Flux rope \\
				\hline
				FLM18 & 21 & 2014/01/02 & (36S,40E) & IF & E & L & -0.50 & right & Flux rope \\
				\hline
				FLM19 & 22 & 2014/01/04 & (26S,28E) & QR &  & L & -1.00 & right & Flux rope \\
				\hline
				FLM11$^{2,4}$ & 23 & 2014/01/04 & (3N,40W) & IF & E & L & -1.00 &  &  \\
				\hline
				FLM17$^{1}$ & 24 & 2014/01/04 & (7S,35W) & IF &  & L & -0.33 & right & Flux rope \\
				\hline
				FLM20 & 25 & 2014/01/04 & (18S,32W) & IF &  & L & -1.00 & right & Flux rope \\
				\hline
				FLM21 & 26 & 2014/01/05 & (22N,13E) & IF & E & R & 0.50 &  &  \\
				\hline
				FLM22 & 27 & 2014/01/06 & (2N,25E) & AR &  & L & -1.00 & left & Sheared Arcade \\
				\hline
				FLM23 & 28 & 2014/01/07 & (29S,38E) & IF & E & L & -0.43 & right & Flux rope \\
				\hline
				FLM19 & 29 & 2014/01/07 & (26S,12W) & IF &  & L & -0.25 & right & Flux rope \\
				\hline
				FLM24 & 30 & 2014/01/07 & (19S,11W) & IF &  & L & -1.00 & right & Flux rope \\
				\hline
				FLM22 & 31 & 2014/01/07 & (3N,12E) & AR &  & L & -1.00 &  &  \\
				\hline
				FLM25 & 32 & 2014/01/08 & (27S,50E) & AR &  & L & -0.50 & right & Flux rope \\
				\hline
				FLM26 & 33 & 2014/01/09 & (22S,10W) & IF &  & L & -1.00 & right & Flux rope \\
				\hline
				FLM27 & 34 & 2014/01/09 & (34S,17W) & IF & E & R & 0.50 &  &  \\
				\hline
				FLM23 & 35 & 2014/01/09 & (29S,15E) & IF & E & L & -0.50 & right & Flux rope \\
				\hline
				FLM25$^{1}$ & 36 & 2014/01/10 & (26S,29E) & AR &  &  & 0.00 &  &  \\
				\hline
				FLM28$^{1}$ & 37 & 2014/01/13 & (12S,32E) & IF & E & R & 0.33 & left & Flux rope \\
				\hline
				FLM29$^{4}$ & 38 & 2014/01/14 & (20N,41W) & IF &  & R & 1.00 &  &  \\
				\hline
				FLM28$^{3}$ & 39 & 2014/01/14 & (15S,28E) & IF & E & R & 1.00 & right & Sheared Arcade \\
				\hline
				FLM28$^{3}$ & 40 & 2014/01/15 & (8S,6E) & IF & E & R & 1.00 & left & Flux rope \\
				\hline
				FLM30 & 41 & 2014/01/15 & (8N,30W) & AR &  & L & -1.00 &  &  \\
				\hline
				FLM31 & 42 & 2014/01/16 & (22S,41E) & QR &  & L & -0.50 & right & Flux rope \\
				\hline
				FLM32$^{5}$ & 43 & 2014/01/18 & (24N,37E) & QR &  & R & 0.50 &  &  \\
				\hline
				FLM33$^{2}$ & 44 & 2014/01/19 & (15S,48E) & AR &  & R & 1.00 & right & Sheared Arcade \\
				\hline
				FLM33$^{2}$ & 45 & 2014/01/20 & (12S,35E) & AR &  & R & 0.5 & right & Sheared Arcade \\
				\hline
				FLM32$^{5}$ & 46 & 2014/01/20 & (25N,15E) & QR &  & R & 1.00 &  &  \\
				\hline
				FLM31$^{1}$ & 47 & 2014/01/20 & (20S,8W) & QR &  & L & -0.33 & right & Flux rope \\
				\hline
				FLM33$^{2}$ & 48 & 2014/01/21 & (18S,26E) & AR &  & L & -1.00 & right & Flux rope \\
				\hline
				FLM31 & 49 & 2014/01/21 & (22S,22W) & AR &  & L & -1.00 & right & Flux rope \\
				\hline
				FLM33$^{2}$ & 50 & 2014/01/22 & (12S,11E) & AR &  & L & -0.50 & right & Flux rope \\
				\hline
				FLM33$^{2}$ & 51 & 2014/01/23 & (12S,1W) & AR &  & L & -0.50 & right & Flux rope \\
				\hline
				FLM33$^{2}$ & 52 & 2014/01/24 & (12S,12E) & AR &  & L & -1.00 & right & Flux rope \\
				\hline
				FLM34 & 53 & 2014/01/24 & (2S,23E) & AR &  & R & 1.00 & left & Flux rope \\
				\hline
				FLM33$^{2}$ & 54 & 2014/01/25 & (10S,29W) & AR &  & L & -1.00 & right & Flux rope \\
				\hline
				FLM35 & 55 & 2014/01/25 & (24S,5E) & IF & E & L & -1.00 & right & Flux rope \\
				\hline
				\multicolumn{10}{c}{\small{$^1$ Noticed low $|C_f|$ value, $^2$ Noticed change in bearing sense, $^3$ Noticed change in skewness, }}\\
				\multicolumn{10}{c}{\small{$^4$ Projection effect and $^5$ Unclear observations}}
			\end{longtable}

			Table~\ref{tab-summary-table} summarizes the hemispheric distribution of the filaments. The 57\% and 43\% of filament observations are in southern hemisphere and northern hemisphere, respectively. Therefore, we find that a significant portion of the filaments (53\%) are left-bearing, while 47\% are right-bearing. Additionally, 84\,\% of filaments with arcade configurations are found in the southern hemisphere. The skewness of the associated coronal arcades reveals a clear hemispheric preference, with all arcades in the southern hemisphere exhibiting rightward skewness and those in the northern hemisphere showing leftward skewness.

			\begin{table}[ht]
				\centering
				\caption{Hemispheric distribution of filaments based on their orientations.}
				\label{tab-summary-table}
				\begin{tabular}{|l|c|c|c|}
					\hline
					\multicolumn{2}{|c|}{\multirow{2}*{Filaments}} & North & South \\
					\cline{3-4}
					\multicolumn{2}{|c|}{} & 13 (43\,\%) & 17 (57\,\%)\\
					\hline
					\multirow{2}*{Bearing Sense} & R & 10 (77\,\%) & 4 (24\,\%)\\
					\cline{2-4}
					& L & 3 (23\,\%) & 13 (76\,\%)\\
					\hline				
				\end{tabular}
			\end{table}

			\section{Summary and Conclusions}
			\label{s-summary} 
			
			We conducted a study on the morphology of several solar filaments observed during the maximum phase of the solar cycle 24. During this period, relatively large number  of filaments appeared on the solar disk. Additionally, we analyzed the properties of the same filaments at different observation times, which increased the reliability of our results. We identified and merged multiple observations of the same filaments, ultimately analyzing a total of 35 filaments from an initial set of 55. These filaments were classified into three categories: quiescent, active region, and intermediate filaments. We tracked each filament to assess whether they underwent eruptions during the observation period. Furthermore, we analyzed the photospheric, chromospheric, and coronal features associated with these filaments to reveal their magnetic structures. We extended our study to investigate the characteristics of filaments exhibiting variability in their observed properties during evolution.

			Using H-alpha intensity observations from the chromosphere, we estimated the chirality of filaments from observed orientations of their barbs. Following the approach of \citet{Pevtsov2003}, we determined the filament chirality by analyzing the bearing sense of filament barbs. We found that all filaments classified as quiescent in this study are following the hemispheric chirality preference. Furthermore, 44\% of active region filaments and 85\% of intermediate filaments from 35 filaments also adhere to this preference, with active region filaments showing a weaker hemispheric preference compared to quiescent filaments, consistent with previous studies \citep{Pevtsov2003}. Overall, 77\% of all filaments analyzed in this study, across all categories, follow the hemispheric chirality preference.  Our result, corroborates earlier studies on filament chirality including those by \citet{Pevtsov2003} and \citet{Hazra2018}. In addition, we evaluated eruptive nature of filaments with different chiralities and differentiated the filament chirality based on the skewness of coronal arcades associated with them \citep{Martin1998}. Our results indicated that all filaments having associated coronal arcades in this study follow hemispheric helicity rule according to their skewness.

			The observed bearing sense of filaments  depends on their magnetic structure, as highlighted earlier \citep{Chen2014,Chen2020}. A dextral filament can exihibit both right- and left- bearing barbs depending on whether the underlying magnetic structure is a flux rope or a sheared arcade,  respectively. We determined magnetic structure of the filaments by combining observations on bearing sense of their barbs and the skewness of coronal arcades above them, as mentioned in \citet{Chen2020}. While skewness data is incomplete for many and right-skewed arcades are more commonly observed when available. Also, we found that all filaments having associated coronal arcades follow hemispheric helicity rule in this analysis. Additionally, the majority of filaments exhibited flux rope magnetic structures as reported  in \citet{Ouyang2017}. A clear hemispheric preference was observed, with right-bearing barbs predominantly in the northern hemisphere and left-bearing barbs in the southern hemisphere, consistent with the findings of  \citet{Pevtsov2003}. This observed relationship between chirality and bearing sense can be attributed to the flux rope morphology detected in most of the filaments in our study. 
			
			The evolution of a helical magnetic flux rope supporting prominence plasma has been described in earlier studies, such as \citet{Ballegooijen1989} and  \citet{DeVore2000}. These studies found that a sheared arcade evolves into a flux rope due to magnetic reconnection. In this study, we detected change in magnetic structure from sheared arcade to flux rope in two filaments during their evolution. These observations validates results of numerical studies. We found that the majority of the filaments in our sample exhibited flux rope morphology. Moreover, we found that all the quiescent filaments in our sample were non-eruptive for long time and followed the flux rope structure, indicating that flux rope morphology may be more stable than the sheared arcade configuration. However, a long lifetime does not mean that the flux rope configuration is less prone to eruption, nor does a short lifetime imply that the sheared arcade configuration is unstable. In other words, the lifetime simply reflects the inherent characteristics of quiescent and active regions. The eruptive behavior of filaments is commonly associated with processes such as magnetic reconnection, kink instability, and flux emergence within the filament plasma structure. In this study, two filaments undergo a morphological transformation from a sheared arcade to a flux rope configuration, highlighting the dynamic evolution of filament magnetic structures.

			The bearing sense of filament barbs and the helicity of the associated active regions have been analyzed in \citet{JainJacob2025a}. That study reports the emergence of positive helicity in the active region associated with one filament. As this filament is located in the southern hemisphere—where positive helicity is typically dominant—this may account for the observed change in barb bearing sense and its evolution into a flux rope configuration. The non-eruptive nature of this filament, together with its transition from a sheared arcade to a flux rope structure, suggests enhanced stability of the flux rope morphology. A similar non-eruptive behavior and morphological transition are observed in the second filament, further supporting this interpretation.

			On the other hand, we identified a few filaments that exhibited changes in their bearing sense and skewness overtime. In this study, we tracked the bearing sense and the skewness of the coronal arcades associated with each filament over an extended period. However, a more detailed study is required to understand the reasons for these changes and to correctly determine the chirality of filaments, as suggested by \citet{Chandra2009}, \citet{Joshi2016}, and \citet{JainJacob2025}. For example, \citet{Chandra2009} observed the emergence of positive helicity magnetic clouds from active regions with negative helicity, which caused changes in the chirality of filaments. Similarly, \citet{Joshi2016} and \citet{JainJacob2025} observed interactions between filaments with different chiralities, which resulted in magnetic restructuring of filaments during evolution.

			Unlike previous statistical studies \citep{Pevtsov2003,Aparna2024}, where filament chirality was determined from a single observation, we tracked the filaments over time and only considered those with consistent chirality.  This approach makes our chirality estimations more reliable compared to earlier studies \citet{Pevtsov2003}, \citet{Aparna2024}, and \citet{Zhou2020}.

			In conclusion, our study provides a detailed analysis of the chirality and magnetic structure of filaments using multiple observations of the same filament over time. This approach allowed for more  accurate determination of their morphology compared to previous studies. Our case study reveals that the majority of filaments exhibit a flux rope magnetic structure, supporting it as the most favorable configuration for solar filaments. Furthermore, we do not find any evidence that filaments deviating from the hemispheric chirality preference are more prone to eruption.

			\begin{acknowledgments}
				The data used in this paper are courtesy of NASA's SDO and AIA science team. This work also utilizes data from GONG and KSO. We thank the anonymous referees for their valuable comments and suggestions which helped us to improve the manuscript. One of the authors, RAM, acknowledges support from the ISRO RESPOND project (No. ISRO/RES/2/437/21‐22).
			\end{acknowledgments}
			
			%

		\end{document}